\documentclass[
 numbers ]
  {aipproc}

\layoutstyle{6x9}
\usepackage{amsmath,amssymb}

\begin{document}

\title{Spontaneous Dimensional Reduction? }
\classification{04.60.-m,04.60.Kz,04.60.Nc}
\keywords      {quantum gravity,dimensional reduction}

\author{Steven Carlip}{
  address={Department of Physics, University of California at Davis, Davis, CA 95616, USA}
}

\begin{abstract}
Over the past few years, evidence has begun to accumulate suggesting that 
spacetime may undergo a ``spontaneous dimensional reduction'' to two 
dimensions near the Planck scale.  I review some of this evidence, and
discuss the (still very speculative) proposal that the underlying mechanism 
may be related to short-distance focusing of light rays by quantum fluctuations.
\end{abstract}

\maketitle


\section{Introduction}

A fundamental goal of quantum gravity is to understand the structure of
spacetime at very short
distances.  This is not easy: while we have a number of interesting ideas and
several promising research programs, a complete, consistent quantum theory 
of gravity remains distant \cite{Carlip1}.  Given these difficulties, one attractive
line of research is to look for places in which different candidates for quantum 
gravity agree.  Even if none of our current models is ultimately correct, such 
areas of agreement suggest deeper underlying structures that might persist 
in the correct theory.  A classic example is black hole thermodynamics: even 
without a full quantum theory of gravity, the Hawking temperature and   
Bekenstein-Hawking entropy are firmly enough established to offer strong 
constraints on any proposed model.

Over the past few years, another area of agreement has started to
emerge.  Hints from several different models of quantum gravity suggest
that at very short distances---perhaps an order of magnitude above the 
Planck scale---spacetime becomes effectively two-dimensional.  Let me
stress that this evidence is far more tentative than the evidence 
for black hole thermodynamics, and may well turn out to be a mirage.  But 
the idea of ``spontaneous dimensional reduction'' is intriguing enough to 
deserve further study.

\section{Hints of dimensional reduction}

I will start by briefly summarizing some of this evidence.  For details, see 
\cite{Carlip2,Carlip3}.

\subsection{Causal dynamical triangulations}

It is well known that the Einstein-Hilbert action of general relativity is
perturbatively nonrenormalizable, and that ordinary methods of
quantum field theory can give us, at best, an effective action that
breaks down above some energy scale \cite{Donoghue,Burgess}.  It
might still be possible, however, to learn about quantum gravity, even  
at arbitrarily high energies, through nonperturbative methods such
as lattice approximations of the path integral.  

The idea of approximating the curved spacetime of general relativity by 
a piecewise flat manifold---a higher dimensional ``geodesic dome''---was 
proposed by Regge more than 50 years ago \cite{Regge}.  One can
put such an approximation on a computer, using Monte Carlo methods 
to generate a ``typical'' ensemble of configurations and numerically 
summing the contributions to the path integral \cite{Williams}.  
Until fairly recently, such attempts generally failed to give a good 
continuum limit; the simulations were instead dominated by ``crumpled'' 
and ``branched polymer'' phases that had little resemblance to the 
spacetime we observe \cite{Loll}.  About a decade ago, though,  Ambj{\o}rn, 
Jurkiewicz, and Loll introduced a new ingredient, a fixed causal structure 
in the form of a prescribed topological time-slicing \cite{AJL,AJL2}.  
The resulting ``causal dynamical triangulations'' (CDT) approach to the 
path integral has proven very promising, not only yielding the correct 
classical de Sitter limit, but also reproducing volume fluctuations 
predicted by quantum minisuperspace models \cite{AJL3,AJL4,Kommu}.  

\begin{figure}
\includegraphics[width=1.95in]{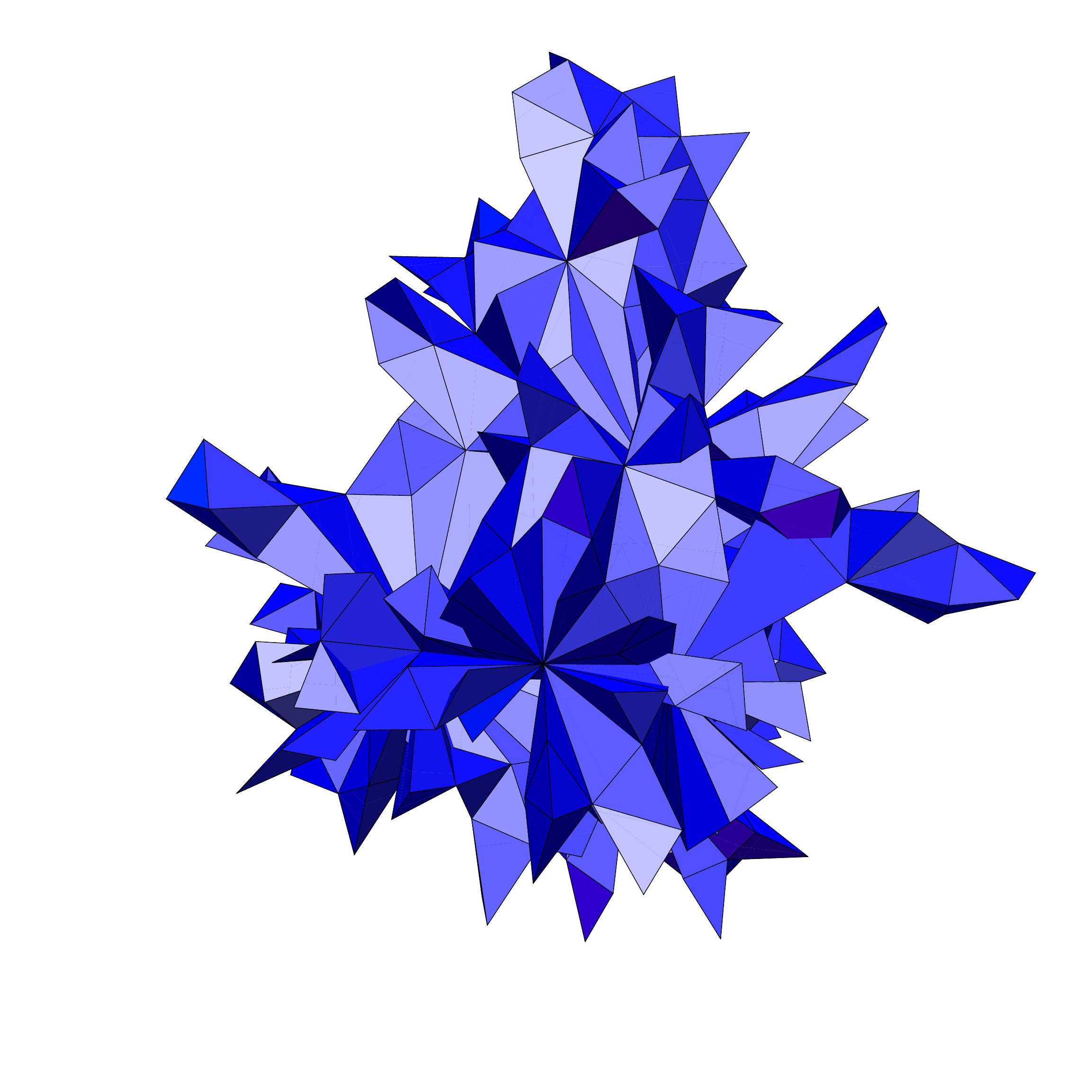}
\includegraphics[width=3.8in]{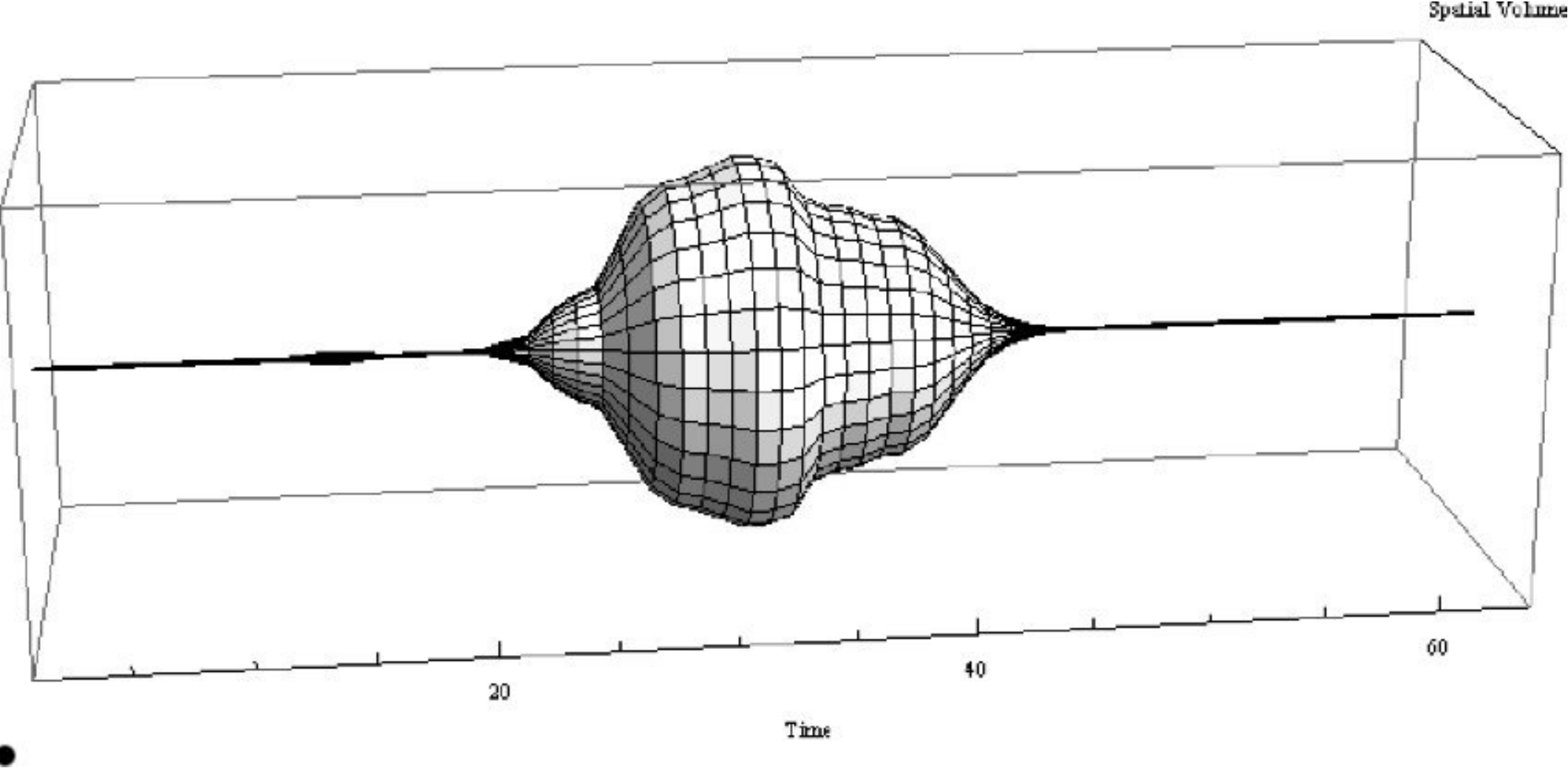} 
\caption{A spatial slice and a typical history contributing to the 
causal dynamical triangulations path integral (R.\ Kommu and M.\ Sachs, 
UC Davis)}
\label{CDT}
\end{figure}

Figure \ref{CDT} illustrates a typical time slice and the volume profile
of a typical history contributing to the path integral, taken from code
developed at UC Davis \cite{Kommu,Sachs}.   It is evident that the 
``paths'' in the path integral are not very smooth.  Moreover, physically 
meaningful quantities are not given by any single configuration, but 
require infinite sums over histories.  As a result, even such basic
observables as the dimension of spacetime become difficult to define
and to extract from the simulations.  

For this last problem, one answer is to use the ``spectral dimension,'' 
the dimension as seen by a random walker \cite{spec}.  The intuitive 
idea is straightforward: a random walker with more dimensions to explore
 will diffuse more slowly from a starting point, and will also take longer 
to return.  Quantitatively, a diffusion process on a $d$-dimensional 
manifold is described by a heat equation
\begin{equation}
\left(\frac{\partial\ }{\partial s} - \Delta_x\right)K(x,x';s) =0 \quad
\hbox{with $K(x,x',0) = \delta(x-x')$} ,
\label{a1}
\end{equation}
with a short distance solution
\begin{equation}
K(x,x';s) \sim (4\pi s)^{-d/2} e^{-\sigma(x,x')/2s}
    \left( 1 + \mathcal{O}(s)\right)
\label{a2}
\end{equation}
where $\sigma(x,x')$ is Synge's world function, essentially the square
of the geodesic distance.  In particular, the return probability $K(x,x,s)$ 
is
\begin{equation}
K(x,x;s) \sim (4\pi s)^{-d/2} .
\label{a3}
\end{equation}
This relationship can now be extrapolated to any space on which a 
diffusion process or random walk can occur.  The spectral dimension is 
then defined as the coefficient corresponding to $d$ in (\ref{a3}).  
While this is only one of several inequivalent definitions of dimension, 
it is a physically important one: the Greens function for a physical field 
can be obtained as a Laplace transform of a heat kernel, so the short 
distance behavior (\ref{a2}) determines physical correlation functions.

The expectation value of the spectral dimension is relatively easy to 
evaluate numerically, since random walks are simple to model.
As Ambj{\o}rn, Jurkiewicz, and Loll first found \cite{AJL3,spec}, and
we have confirmed \cite{Kommu}, the CDT spectral dimension is four
for ``long'' random walks, as required for the emergence of a 
good classical limit, but falls to two for ``short'' random walks.
The corresponding Greens functions are those of four-dimensional
fields at large scales, but those of two-dimensional fields at small
scales.  This phenomenon, which occurs at scales of about fifteen Planck 
lengths \cite{Cooperman}, is perhaps the strongest piece of evidence 
for spontaneous dimensional reduction.

\subsection{Exact renormalization group}

Further evidence for short distance dimensional reduction comes
from the analysis of the renormalization group flow for general 
relativity \cite{Reuter,Nieder,Litim}.  This research program was inspired 
by Weinberg's suggestion that gravity might be ``asymptotically safe''
\cite{Weinberg}: that is, even though the theory is nonrenormalizable,
it might have an ultraviolet fixed point at which all coupling constants 
are finite and well-behaved.  If, in addition,  the space of such UV 
fixed points (the ``critical surface'') turns out to be finite dimensional, 
the IR coupling  constants would be determined by a finite number of 
UV parameters:  not quite renormalizability, but almost as good.

This possibility has been investigated by truncating the gravitational
effective action to a relatively small number of terms and using
exact renormalization group techniques.  While agreement is not
universal, the evidence points to the existence
of an ultraviolet fixed point, whose qualitative behavior does not change 
as more higher derivative terms and matter interactions are added back
in to the effective action.  (See \cite{Litim} for a review of the current
status.)  For our purposes, the most interesting feature is that operators
acquire large anomalous dimensions---quantum corrections to their
naive ``engineering dimensions''---at the UV fixed point.  As a result,
they behave very much as if they lived in a two-dimensional spacetime.  
Greens functions, for instance, for both matter and the 
gravitational field itself, look just as one would anticipate
from the CDT results: effectively four-dimensional in the IR but
two-dimensional in the UV.  Moreover, a calculation of the renormalization
group behavior of the spectral dimension seems to reproduce the CDT 
results \cite{Reuter2}.

In retrospect, the appearance of two-dimensional behavior in this 
context is not too surprising.  Two is the unique dimension in which the 
Einstein-Hilbert action becomes dimensionless, and there are 
fairly general arguments that if an ultraviolet fixed point exists, it must 
look effectively two-dimensional \cite{Percacci}.  We thus have another 
hint of dimensional reduction.

\subsection{Further evidence}

Further indications of spontaneous dimensional reduction at short 
distances come from a number of other approaches to quantum gravity:
\begin{itemize}\renewcommand{\labelitemi}{\labelitemii}
\item An old string theory result \cite{Atick} is that at a very high 
temperature $T$, the free energy of a gas of strings in a  
volume $V$ behaves as $F/VT\sim T$.  This may be recognized as 
the expression for the free energy of a two-dimensional field theory; 
in fact, Atick and Witten suggested that high-temperature string 
theory might behave like ``a lattice theory with a (1+1)-dimensional 
field theory on each lattice site.''
\item As Modesto has noted \cite{Modesto}, the area spectrum of loop
quantum gravity looks as if elementary spatial areas drop from being
two-dimensional at large scales to one-dimensional at small scales.
\item In Ho{\v r}ava-Lifshitz models, modifications of general relativity
in which power-counting renormalizability is restored at the expense of
breaking Lorentz invariance, the spectral dimension also drops to two
at small distances \cite{Horava}.
\item The Myrheim-Meyer dimension for a random causal set is 
approximately $2.38$ \cite{Sorkin}.
\end{itemize}
There are also indications from noncommutative geometry \cite{Connes}
and fractal approaches to high energy physics \cite{Calcagni} that could 
point toward dimensional reduction; see also \cite{Anchordoqui}.

\section{The short distance Wheeler-DeWitt equation}

Let me now turn to another piece of evidence for short distance
dimensional reduction, which will be the focus of the remainder of
this article: the ``strong coupling'' behavior of the Wheeler-DeWitt
equation.  The Wheeler-DeWitt equation---essentially the quantum
version of the Hamiltonian constraint of general relativity---is usually
written in units $G=\hbar=1$.  If we instead retain the coupling 
constants, it takes the form
\begin{equation}
\left\{ 16\pi\ell_p^2G_{ijkl}\frac{\delta\ }{\delta g_{ij}} \frac{\delta\ }{\delta g_{kl}}
    - \frac{1}{16\pi\ell_p^2}\sqrt{g}\,{}^{(3)}\!R\right\}\Psi[g] = 0 ,
\label{b1}
\end{equation}
where 
$ 
G_{ijkl} = \frac{1}{2}g^{-1/2}\left( g_{ik}g_{jl} + g_{il}g_{jk} -
  g_{ij}g_{kl}\right)
$
is the DeWitt metric on the space of metrics and $\ell_p= \sqrt{\hbar G}$ 
is the Planck length.  The short distance/strong coupling limit is reached 
by taking $\ell_p\rightarrow\infty$.\footnote{Since $\ell_p $ is a 
dimensionful quantity, we must be a bit careful of what this limit means.  
In general, solutions of (\ref{b1}) have support on metrics with variations 
at arbitrary scales.  Taking $\ell_p$ large really means focusing  in on 
the dependence on metrics that vary rapidly at the Planck scale.}   This 
is also the ``ultralocal'' limit \cite{Isham}: spatial derivatives in the
Wheeler-DeWitt equation only occur in the scalar curvature term, so 
as $\ell_p$ increases the coupling between neighboring points becomes 
weak. 

The behavior of the Wheeler-DeWitt equation in this limit was studied
extensively in the 1980s \cite{HPT,Teitelboim,Pilati2,Rovelli,Husain}.
The key features are already captured in the classical behavior.  In the
infinite $\ell_p$ limit, spatial points decouple, and the classical solution
is a collection of Kasner spaces---
\begin{multline}
ds^2 = dt^2 - t^{2p_1}dx^2 - t^{2p_2}dy^2 - t^{2p_3}dz^2 \\
\hbox{ with\ \ $-\frac{1}{3}<p_1<0<p_2<p_3,\quad
p_1+p_2+p_3 = 1 = p_1^2+p_2^2+p_3^2$,\ \ } 
\label{b3}
\end{multline}
with independent parameters $p_i$ and independent axes at each point.  
If one now treats the scalar curvature as a perturbation, one obtains 
BKL (Belinskii-Khalatnikov-Lifshitz) behavior \cite{BKL}: the metric at 
any given point spends most of its time in a Kasner form, with 
neighboring points now weakly coupled, but it occasionally undergoes 
a rapid, chaotic ``bounce'' to a new Kasner space with different axes 
and exponents.  The result is essentially a random, fluctuating collection
of Kasner spaces, with a known probability distribution \cite{Kirillov2}.

To connect this behavior to dimensional reduction, observe that while
Kasner space is (3+1)-dimensional, it acts in important ways as
if it were (1+1)-dimensional \cite{Carlip2,Carlip3,Chicone}.  For 
example, consider a geodesic starting at a random point with a 
random initial direction.  It is not hard to show that for almost every 
such geodesic, the proper distance traversed along the path will be 
almost entirely in one spatial dimension, along the $p_1$ axis.  In a 
sense, an observer probes only one dimension of space.  Similarly, 
as $t\rightarrow0$, the particle horizon in Kasner space becomes 
cigar-shaped, stretching out along the $p_1$ axis while becoming  
tiny in the orthogonal directions.

Since the spectral dimension measures the behavior of random walks,
one might expect this aspect of Kasner space---an ``effective infrared 
dimension'' that differs from four \cite{Hu}---to appear in the heat
kernel.  As far as I know, the heat kernel for Kasner space has not been
computed exactly, but two different approximations \cite{Futamase,Berkin}
both give behavior of the form
\begin{equation}
K(x,x;s) \sim \frac{1}{4\pi s^2}\left[ 1 + \frac{a}{t^2}\,s + \dots \right] .
\label{b4}
\end{equation}
For small enough $t$, the $1/s$ term will dominate; we can see from
(\ref{a2}) that this is characteristic of two-dimensional behavior.  A 
similar result can be obtained from the Seeley-DeWitt expansion of the 
heat kernel \cite{Carlip3}.

We are thus led to a tentative picture of small scale spacetime as a 
collection of chaotic, rapidly varying, weakly coupled ``nearly 
(1+1)-dimensional'' spaces.  One must worry about experimental
implications of the resulting violation of Lorentz invariance.  But this
violation is ``nonsystematic''---its parameters vary rapidly in space 
and time---and such phenomena seem much harder to detect than
``systematic'' violations \cite{Basu}.

\section{Short Distance Asymptotic Silence?}

The BKL behavior of the preceding section has been carefully studied 
in a different context, that of classical cosmology near a generic spacelike
 singularity \cite{HUR}.  There, the essential feature is what is called 
``asymptotic silence.''  Light cones along world lines orthogonal to the
singularity collapse to lines as the singularity is approached, leading 
to causal disconnection of nearby points.  This behavior can be described
as an anti-Newtonian limit: rather than approaching infinity, as in the
Newtonian limit, the effective speed of light shrinks to zero.  If this 
causal disconnection occurs rapidly enough, the structure of the field 
equations leads almost automatically to BKL behavior.

The ultralocal nature of the short distance Wheeler-DeWitt equation 
provides one indication of asymptotic silence in quantum gravity.  Let
us search for others.  Note that unlike the cosmological setting, in 
which the initial singularity defines a preferred set of world lines, we 
now need light cones to collapse along world lines orthogonal to an 
\emph{arbitrary} initial surface.

The behavior of congruence of null geodesics---a pencil of light---is 
governed by the Raychaudhuri equation \cite{Raychaudhuri,Poisson},
\begin{equation}
\frac{d\theta}{d\lambda} = -\frac{1}{2}\theta^2 - \sigma_a{}^b \sigma_b{}^a
     + \omega_{ab}\omega^{ab} - 16\pi G T_{ab}k^a k^b ,
\label{c1}
\end{equation}
where $\theta$ is the expansion---essentially the logarithmic derivative
of the cross-sectional area of the bundle---and $\sigma$ and $\omega$
are the shear and vorticity.  Ideally, we would like to understand the
quantum version of this equation, but this would require a much deeper
knowledge of quantum gravity than we have so far.  If the phenomena we 
are interested in take place a bit above the Planck scale, though, we might
be able to learn something from the semiclassical version of (\ref{c1}).

Semiclassically, the expansion and shear terms on the right-hand side of
(\ref{c1}) are nonnegative, and focus geodesics.\footnote{This might no
longer be the case in the full quantum theory, since the regularized value 
of a positive quantity can be negative, but we do not yet know enough
 about quantum gravity to say more.}  
The vorticity term defocuses geodesics, but if no vorticity is initially 
present, none will be generated.  The behavior of the stress-energy 
term is thus key.   We know that for many forms of matter, most 
vacuum fluctuations have negative energy, and thus defocus geodesics 
\cite{Fewster,Fewster2}.  But these negative energy fluctuations 
have a strict lower bound, while the rarer positive fluctuations are 
unbounded.  The question, then, is whether the behavior of light 
cones near the Planck scale is dominated by the frequent negative 
fluctuations or the rare but potentially very large positive fluctuations.

In \cite{Pitelli}, Mosna, Pitelli and I investigated this question in the
simpler setting of two-dimensional dilaton gravity.  The advantage
of this lower-dimensional model is that the full probability distribution
for vacuum stress-energy fluctuations is known \emph{exactly}
\cite{Fewster}, allowing a much more reliable calculation.  We found
that the rare positive energy fluctuations win: the expansion collapses
to $-\infty$ in a characteristic time of about fifteen Planck times.
Intriguingly, this is the same scale at which the spectral dimension
falls to two in causal dynamical triangulations.

This result may be understood qualitatively as follows.  In probability
theory, there is a famous result known as ``gambler's ruin'': in
the long run, a gambler with finite wealth playing against a house with 
infinite resources will always lose.  In the case at hand, negative energy
fluctuations are a finite resource, while positive energy fluctuations 
are unbounded.  In the short run, the frequent negative fluctuations 
will tend to defocus geodesics, but a large positive fluctuation 
can drive the expansion negative enough that the nonlinearities in 
(\ref{c1}) dominate.  Once this happens, the system``goes broke,'' 
and no subsequent negative fluctuation can compensate.

\begin{figure}
\includegraphics[width=5in]{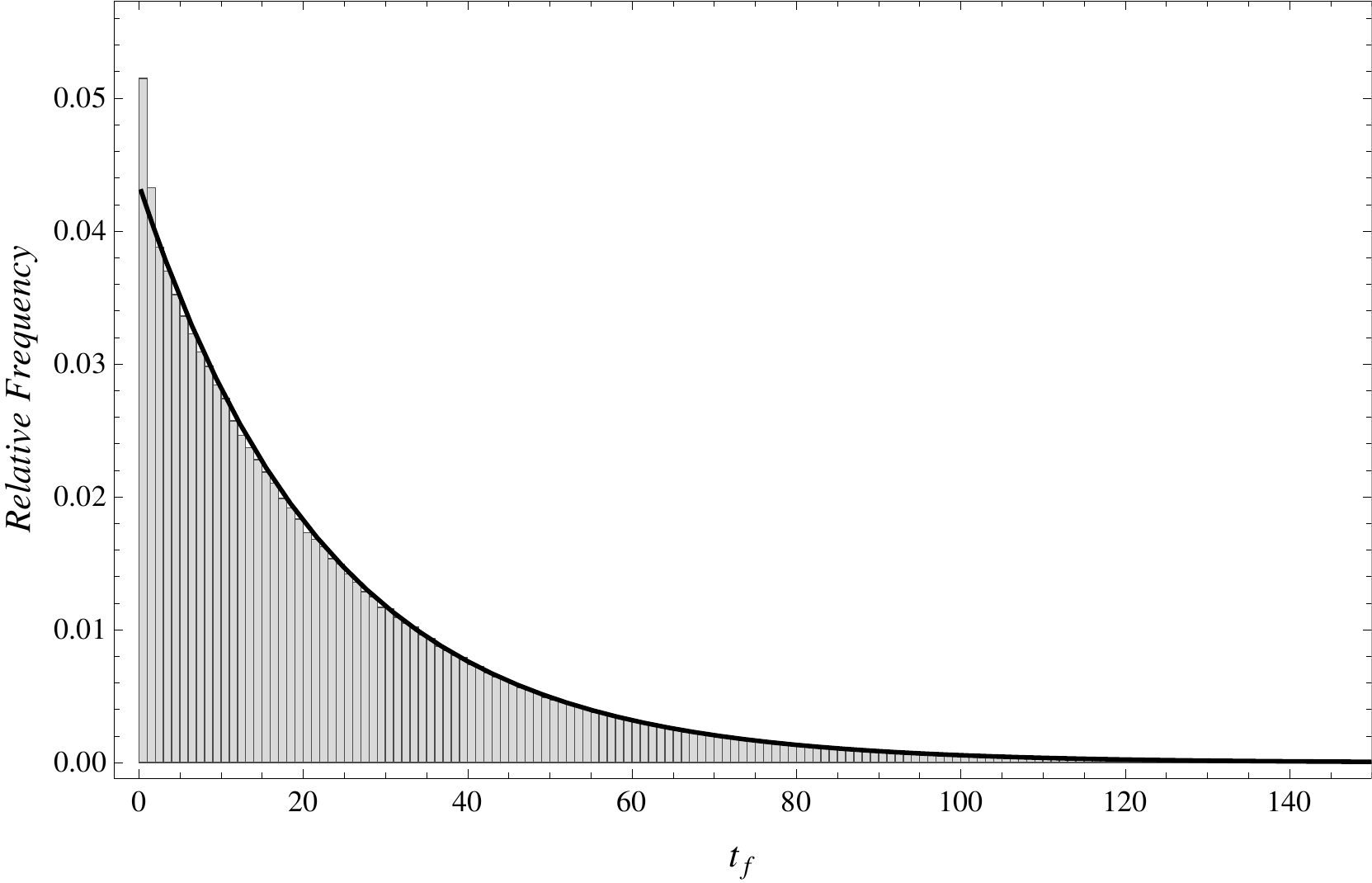}
\caption{Probability of the expansion diverging to $-\infty$ as a
function of Planck time steps.  The solid line is the exponential
distribution (\ref{c2}).}
\label{flucs}
\end{figure}

More concretely, consider a discrete model in which fluctuations
take place over a ``smearing time'' $\Delta t$, which we  take to
be the Planck time.  Suppose the minimum value of 
a stress-energy fluctuation smeared over this time interval is 
$\min(T_{ab}k^ak^b) = -\mathcal{T}$.  Note that if the expansion 
falls below the critical value ${\bar\theta} =-\sqrt{32\pi G\mathcal{T}}$, 
the right-hand side of ({\ref{c1}) will be strictly negative, and $\theta$ 
will drop rapidly and irreversibly.   

Now begin with an arbitrary value of the expansion.  Initially, the 
relatively frequent negative energy fluctuations will tend to push 
$\theta$ to a value on the order of $|{\bar\theta}|$, at which point 
the nonlinearities will prevent it from growing further.  Let $\rho$ 
be the probability of a positive vacuum fluctuation with 
a value greater than $2\mathcal{T}$.  Treating the fluctuations
as a Poisson process,\footnote{This is not strictly accurate, since the
fluctuations are not uncorrelated, but the correlation function drops 
off very quickly with time \cite{Pitelli}.} it is easy to show that the
probability for $\theta$ to be pushed below the critical value
$\bar\theta$ in a time $n\Delta t$ is approximated by an 
exponential distribution
\begin{equation}
\mathit{Prob}(t_{\mathit{collapse}}=n\Delta t) \sim  e^{-\rho t/\Delta t} .
\label{c2}
\end{equation}
Figure \ref{flucs} shows the result of 10 million computer runs simulating
the Raychaudhuri equation, each using a random sequence of stress-energy 
fluctuations taken from the exact probability distribution found in 
\cite{Fewster}.  The estimate (\ref{c2}) fits surprisingly well.

So far, these results are limited to 1+1 dimensions.  Recently, 
though, the probability distribution for vacuum stress-energy fluctuations 
has been computed in 3+1 dimensions \cite{Fewster2}.  The full analytic 
form of the distribution is not known, and it is not clear whether a 
simulation like the one that led to figure \ref{flucs} will be feasible.  
But the large positive energy tail is under good control, and it should 
be possible to obtain at least an estimate analogous to (\ref{c2}).  It
may also be possible to directly study the shapes of light cones in 
causal dynamical triangulations to test the appearance of asymptotic 
silence.

\section{Where We Stand}

The analysis of the preceding section is suggestive, but not conclusive.  In
particular, we do not understand the full quantum Raychaudhuri equation, 
and we don't know whether the classical analysis of asymptotic silence
really carries over to the quantum case.  Still, we are left with an intriguing
picture of the small scale structure of spacetime:
\begin{itemize}\renewcommand{\labelitemi}{\labelitemii}
\item At distances of a few times the Planck scale, quantum fluctuations
cause light cones to collapse, resulting in a causal disconnection of nearby 
points in spacetime.
\item The dynamics of general relativity, either classical or quantum, then 
leads to BKL behavior: short distance spacetime looks like a nearly random, 
weakly coupled, chaotically changing collection of Kasner space domains.
\item The effective two-dimensional behavior of Kasner space, in which
the dynamics is concentrated  along a preferred direction, can be
interpreted as a sort of spontaneous dimensional reduction of spacetime.
\item Lorentz violations occur near the Planck scale, but these are 
nonsystematic, and average out at larger scales.
\end{itemize}

If this picture is correct, it suggests a new approach to quantizing 
gravity. In a very different context, several authors have investigated 
a formulation of general relativity in a setting in which two pairs of 
dimensions have very different characteristic scales \cite{tHooft,Verlinde,Kabat}.
It may be possible to adapt these methods to the picture described here.

In particular, suppose we can write the metric locally in the form
\begin{equation}
ds^2 = \ell_\parallel^2g_{\mu\nu}dx^\mu dx^\nu 
         +\ell_\perp^2h_{ij}dx^idx^j
\label{d1}
\end{equation}
and that ``transverse'' derivatives $\partial_i$ are small.  Then the
Einstein-Hilbert action is approximately
\begin{equation}
I \sim \frac{\ell_\perp^2}{\ell_p^2}\int d^2x d^2y \sqrt{h}
   \left( \sqrt{g}R_g 
   + \frac{1}{4}\sqrt{g}g^{\mu\nu}\partial_\mu h_{ij}\partial_\nu h_{kl}
    \epsilon^{ik}\epsilon^{jl}\right) ,
\label{d2}
\end{equation}
an expression that looks very much like a two-dimensional action for the 
transverse metric $h_{ij}$.  

This action is not quite conformally invariant---the trace of the effective 
stress-energy tensor is $T \sim \Box \sqrt{h}$.  But the deviation can be 
small.  In Kasner space, for example, $\sqrt{h}$ goes to zero for small $t$.  
It may thus be possible to use powerful techniques from two-dimensional 
conformal field theory to analyze this system.  Indeed, borrowing ideas
from \cite{Solodukhin}, we can set $h_{ij} = (1+\varphi)\sigma_{ij}$
with $\det\sigma_{ij}=1$, and find that to lowest order, $\varphi$ is
a Liouville field with a central charge of order $A_\perp/\ell_p^2$.

Work on this program is just starting, and it remains to be seen 
how fruitful it will be.  In particular, the eikonal methods of 
\cite{tHooft,Verlinde,Kabat} were designed for a fixed, global splitting
of the metric into ``longitudinal'' and ``transverse'' directions, and it
will take a good deal of work to understand how to connect neighboring
domains with different preferred directions.  
But these results suggest a new and interesting possibility for quantum 
gravity, and some promising avenues for future study.


\begin{theacknowledgments}
Portions of this project were carried out at the Peyresq 15 Physics Conference 
with the support of OLAM Association pour la Recherche Fundamentale, 
Bruxelles.  This work was supported by the U.S.\ Department of Energy under grant
DE-FG02-91ER40674.
\end{theacknowledgments}

\end{document}